# The de Hass-van Alphen quantum oscillations in BaSn$_3$ superconductor with multiple Dirac fermions

Gaoning Zhang[1,†], Xianbiao Shi[2,3,†], Xiaolei Liu[1,†], Wei Xia[1], Hao Su[1], Leiming Chen[4], Xia Wang[1,5], Na Yu[1,5], Zhiqiang Zou[1,5], Weiwei Zhao[2,3*], Yanfeng Guo[1*]

[1] School of Physical Science and Technology, ShanghaiTech University, Shanghai 201210, China
[2] State Key Laboratory of Advanced Welding & Joining and Flexible Printed Electronics Technology Center, Harbin Institute of Technology, Shenzhen 518055, China
[3] Key Laboratory of Micro-systems and Micro-structures Manufacturing of Ministry of Education, Harbin Institute of Technology, Harbin 150001, China
[4] School of materials science and engineering, Henan key laboratory of aeronautic materials and application technology, Zhengzhou University of Aeronautics, Zhengzhou, Henan, 450046
[5] Analytical Instrumentation Center, School of Physical Science and Technology, ShanghaiTech University, Shanghai 201210, China

By measuring the de Hass-van Alphen effect and calculating the electronic band structure, we have investigated the bulk Fermi surface of the BaSn$_3$ superconductor with a transition temperature of ~ 4.4 K. Striking de Haas-van Alphen (dHvA) quantum oscillations are observed when the magnetic field $B$ is perpendicular to both (100) and (001) planes. Our analysis unveiled nontrivial Berry phase imposed in the quantum oscillations when $B \perp$ (100), with two fundamental frequencies at 31.5 T and 306.7 T, which likely arise from two corresponding hole pockets of the bands forming a type-II Dirac point. The results are supported by the *ab initio* calculations indicating a type-II Dirac point setting and tilting along the high symmetric K-H line of the Brillouin zone, about 0.13 eV above the Fermi level. Moreover, the calculations also revealed other two type-I Dirac points on the high symmetric Γ-A direction, but slightly far below the Fermi level. The results demonstrate BaSn$_3$ as an excellent platform for the study of not only exotic properties of different types of Dirac fermions in a single material, but also the interplay between nontrivial topological states and superconductivity.

[†]The authors contributed equally to this work.

[*]Corresponding authors:
wzhao@hit.edu.cn,
guoyf@shanghaitech.edu.cn.

## I. INTRODUCTION

The nontrivial topological states in solids have brushed up our knowledge on band theory, which provide a platform to conveniently study the intriguing physics of some elementary particles with properties akin to those theoretically predicted in high-energy physics, such as Dirac fermions [1-3], Weyl fermions [4-7], Majorana fermions [8-14], and even other exotic new fermions that have no counterparts in high-energy physics [15-19]. The realization of these particles is owing to the numerous symmetries of solids and their versatile operations, which whereas is somewhat difficult in high-energy physics and most of the particles consequently have not been discovered yet. With the protection by both time-reversal and space-inversion symmetries, the massless Dirac fermions can appear as low-energy quasiparticles in the Dirac semimetals (DSMs), such as $Na_3Bi$ [2, 3] and $Cd_3As_2$ [20, 21], around the Dirac points (DPs) which are formed by the four-fold degenerate linear bands crossing. Once the time-reversal symmetry or space-inversion symmetry is broken, a DP will split into two Weyl points (WPs) that are characterized by a pair of Weyl fermions with opposite chiralities, such as in the Weyl semimetals (WSMs) TaAs series [4-7]. In the momentum space, both Dirac and Weyl cones can be tilted via modulating the tilting term in the Hamiltonian. Unlike the case in high-energy physics, where the Lorentz invariance is strictly required but not necessary in condensed matter physics, which could be broken by additional momentum dependent term, thus forming the type-II DSMs and WSMs, such as the $PtTe_2$ [22-26] and the $MoTe_2$ families [27-31]. The type-II DPs and WPs are still symmetry-protected while the spectra of the Dirac and Weyl cones are strongly tilted. Different from the type-I Dirac or Weyl fermions, the type-II Dirac or Weyl fermions are generally located at the boundary between the electron-like and hole-like pockets of the Brillouin zone (BZ), which are expected to display unique properties, such as Klein tunneling in momentum space [32], angle-dependent chiral anomaly [33], modified anomalous Hall conductivity [34], and topological Lifshitz transition [35], etc. In light of the different physical properties of type-I and type-II Dirac or Weyl fermions, the combination of them in a single material is rather attractive for the realization of multiple functionalities. However, it is definitely challenging because of the different symmetry requirements for them. The coexistence of different nontrivial topological states in a single material is expected to produce exotic physical properties, which has been reported in some compounds [15-19]. Regarding the combination of different types of Dirac fermions in a single material, the ternary wurtzite CaAgBi family of materials was predicted to be excellent targets, which however is still waiting for experimental verification [36].

Recently, the exploration of nontrivial topological states in superconductors has garnered increasing attentions [8-14], because superconductors hosting nontrivial topological states are thought to be a good place to discovery Majorana Fermion. The interplay of nontrivial band topology and superconductivity might give rise to topological superconductivity with spin-momentum locking of topological surface states hosting the long-pursuing Majorana Fermion, which is potential for the use in low decoherence topological quantum computation [37-42]. $LnM_3$ (Ln = Y and rare-earth elements, M = Pb, TI, In, Ga, and Sn) compounds with the $AuCu_3$-type structure cover an array of interesting physical properties, such as superconductivity with relatively high superconducting transition temperature $T_c$, heavy fermion behavior, quantum critical points [43, 44], etc. Due to ionic radius mismatch between the alkaline earth metal ($A$) and Sn ions, $ASn_3$ generally crystallize in the variant structures of $AuCu_3$. For example, $CaSn_3$ is in the $AuCu_3$ structure [45], while $BaSn_3$ ($T_c$ = 4.3 K) has a hexagonal structure and $SrSn_3$ ($T_c$ = 5.4 K) crystallizes in a rhombohedral structure [46]. Up to now, the physical properties of $ASn_3$ have been less studied as compared with $LnM_3$. The recent studies on $CaSn_3$ have revealed it as a weakly coupled type-II superconductor hosting eight pairs of WPs in the BZ with closed loops of surface Fermi arcs [47-49]. In light of the intriguing properties of $CaSn_3$, investigation on the analogue $BaSn_3$ in terms of the electronic structure to examine possible nontrivial topological states is very valuable. By performing de Hass-van Alphen (dHvA) effect measurements and *ab initio* calculations, we have demonstrated herein that the $BaSn_3$ superconductor simultaneously hosts type-I and type-II Dirac fermions.

## II. EXPERIMENTAL

High quality BaSn$_3$ crystals were grown from a self-flux method. The Ba pieces and Sn lumps were mixed in the molar ratio 1 : 6 and sealed in an evacuated quartz tube under the vacuum of about $10^{-4}$ Pa. The tube was slowly heated to 700 °C in 15 h, kept at this temperature for 10 h in order to mix the solution uniformly, and then slowly cooled down to 400 °C at a rate of 3 °C /h. The single crystals were obtained by immediately putting the tube in a high speed centrifugation at this temperature to remove the excess Sn. Columnar hexagonal shaped crystals were obtained after breaking the tube in the glove box protected by high-purity Argon gas. A typical crystal of BaSn$_3$ is shown on a millimeter grid by the inset of Fig. 1(c). The crystallographic phase and crystal quality were examined on a Bruker D8 single crystal X-ray diffractometer (SXRD) with Mo $K_\alpha$ ($\lambda$ = 0.71073 Å) at 300 K. The diffraction pattern could be nicely indexed in the space group $P6_3/mmc$ (No.194) with lattice parameters $a$ = $b$ = 7.2424 Å, $c$ = 5.4870 Å, and $\alpha$ = $\beta$ = 90°, $\gamma$ = 120°, consistent with previously reported [46]. A schematic view of the crystal structure in Fig. 1(a) is drawn based on the refinement results from the SXRD data. The chemical compositions and uniformity of stoichiometry were checked by the energy dispersive spectroscopy (EDS) at several spots on the crystals used for later measurements. The results are shown in Fig. 1(c), revealing a good stoichiometry. The perfect reciprocal space lattice of SXRD without any other miscellaneous points, seen in Figs. 1(d) - (f), indicates pure phase and high quality of the crystal. Magnetic susceptibilities measured at the zero-field-cooling (ZFC) and field-cooling (FC) modes with the magnetic field $B$ of 10 Oe with $B \perp$ (001) plane. The electrical transport and magnetic measurements were respectively performed on a physical property measurement system (PPMS) and a magnetic property measurement system (MPMS) from Quantum Design. Due to sensitivity of the crystal to air, a layer of paraffin was coated on the crystal to isolate the air for magnetization measurements.

The first-principles calculations were carried out within the framework of the projector augmented wave (PAW) method [50, 51], by employing the generalized gradient approximation (GGA) [52] with Perdew-Burke-Ernzerhof (PBE) [53] formula, as implemented in the Vienna *ab initio* Simulation Package (VASP) [54-56]. A kinetic energy cutoff of 500 eV and a Γ-centred $k$ mesh of 6×6×9 were utilized in all calculations. The energy and force difference criterion were defined as $10^{-6}$ eV and 0.01 eV/Å for self-consistent convergence and structural relaxation, respectively. The WANNIER90 package [57-59] was adopted to construct Wannier functions from the first-principles results without an iterative maximal-localization procedure. The WANNIERTOOLS [60] code was used to investigate the topological features of surface state spectra.

## III. RESULTS AND DISCUSSION

The magnetic susceptibility of BaSn$_3$ crystal is presented in Fig. 1(b) with the inset showing the temperature dependence of longitudinal resistivity $\rho_{xx}$ measured with the electrical current $I$ = 50 μA along the $c$-axis. Both of them display a superconducting onset temperature $T_c^{onset}$ ~ 4.4 K, which are rather close to the values reported in other literatures [46]. The $B$ dependent magnetization at 2 K is shown by the right inset of Fig. 1(b), exhibiting clear hysteresis that points to a type-II superconductivity of BaSn$_3$. Since the element Sn has a $T_c$ of about 3.6 K with a type-I superconductor nature, the different superconductivity of BaSn$_3$ and Sn excludes the possibility that the superconductivity of BaSn$_3$ is from the Sn flux rather than from the bulk.

Seen in Figs. 2(a) and (d), clear dHvA quantum oscillations in the isothermal magnetizations starting from $B$ = 2 T in the temperature range of 2 - 15 K are visible when $B$ is perpendicular to both (001) and (100) planes, likely indicating anisotropic Fermi surface (FS) of BaSn$_3$. In Figs. 2(a) and (d), we present the oscillatory components of magnetization for BaSn$_3$ at different temperatures, which were obtained after subtracting the background, i.e. $\Delta M = M - M_{background}$, the similar method widely used in analyzing the dHvA quantum oscillations. In general, the dHvA quantum oscillations could be well described by the Lifshitz-Kosevich (L-K) formula [61]:

$$\Delta M \propto -B^\lambda R_T R_D R_s \sin[2\pi\left(\frac{F}{B} - \gamma - \delta\right)],$$

where $R_T = \alpha T\mu/B\sinh(\alpha T\mu/B)$ with $\mu$ being the ratio of effective cyclotron mass $m^*$ to free electron mass $m_0$ and $\alpha = (2\pi^2 k_B m_0)/(\hbar e)$, $R_D = \exp(-\alpha T_D \mu/B)$ with $T_D$ being the sample dependent Dingle temperature, and $R_S = \cos(\pi g \mu/2)$. The exponent $\lambda$ is determined by the dimensionality, where $\lambda = 1/2$ and 0 are for the three-dimensional (3D) and 2D cases, respectively. The oscillation part is described by the sine term with a phase factor $\gamma - \delta$, in which $\gamma = \frac{1}{2} - \frac{\phi_B}{2\pi}$ with $\phi_B$ being the Berry phase. The phase shift is determined by the dimensionality of the FS and takes a value of 0 for 2D and ±1/8 ( "-" for electronlike pocket and "+" for the holelike pocket) for 3D case [62]. The fast Fourier transform (FFT) analysis of dHvA oscillations are depicted in Figs. 2(b) and (e) after first order differential of the $M$ vs. $1/B$. A single frequency labeled as $\chi = 300.8$ T was obtained from the analysis on the oscillations when $B \perp (001)$, corresponding to the calculated Band 3 in Fig. 3 (b) assigned by a carful comparison. Multiple frequencies associated with the oscillations when $B \perp (100)$ were obtained in the FFT spectra as shown in Fig. 2(e). In order to achieve more accurate fits, we separated the low and high frequency components with band-pass filtering and fitted them individually. The multiple frequencies in the FFT spectra indicate the presence of multiple Fermi pockets around the Fermi energy level $E_F$. Moreover, the anisotropic dHvA oscillations confirm the 3D nature of FS in $BaSn_3$. From the temperature dependent FFT spectra, four main frequencies for $B \perp (100)$ could be identified, labeled as α (31.5 T), α* (126T), $\beta_1$ (196.6T), $\beta_2$ (306.7T). Here α* = 4α is the harmonic value of α. The angle dependent frequencies α, $\beta_1$, and $\beta_2$ are comparable to those calculated for the Band 1 and Band 2, respectively, as shown in Fig. 3(b). The area of FS can be determined by the Onsager relation $F = A_F (\varphi/2\pi^2)$, where $\varphi = h/e$ is the magnetic flux and $A_F$ is the FS area. The derived parameters are summarized in Table I.

From the L-K formula fitting, the effective mass $m^*$ can be obtained through the fit of the temperature dependent oscillation amplitude to the thermal damping factor $R_T$, as shown by the insets of Figs. 2(b) and (e). For all obtained oscillation frequencies, the effective masses are estimated as in the range of $0.135m_0$ - $0.29m_0$, which are very close to the DFT calculated values in the range of $0.1m_0$ - $0.3m_0$. Fitting to the field dependent amplitudes of the quantum oscillations at 2 K, shown in Figs. 2(c) and (f), gives the Dingle temperatures $T_D$ = 12 K and 18 K for band $\alpha$ and $\chi$, respectively.

To achieve in-depth insights into the dHvA oscillations, the Landau level (LL) index fan diagram is constructed, aiming to examine the Berry phase of $BaSn_3$ accumulated along the cyclotron orbit [62]. The LL index phase diagram of $\alpha$ and $\chi$ are shown in Figs. 2(c) and (f), respectively, in which the valley positions of $dM/dB$ against $1/B$ were assigned to be integer indices and the peak positions of $dM/dB$ were assigned to be half-integer indices. A good linear fitting gives the intercepts of 0.4, -0.04, -0.35 and 0.08 corresponding to $\alpha$, $\beta_1$, $\beta_2$ and $\chi$, respectively. The Berry phases $\phi_B$ derived from our analyses are $(0.8+0.25)\pi$, $(-0.08+0.25)\pi$, $(-0.7+0.25)\pi$ and $(0.16-0.25)\pi$ for the hole pockets $\alpha$, $\beta_1$, $\beta_2$ and the electronic pocket $\chi$. The Berry phases for the $\alpha$ and $\beta_2$ bands are close to $\pi$, implying the nontrivial topology of these two hole bands in $BaSn_3$, while the hole band $\beta_1$ and electronic band $\chi$ are topologically trivial. All derived parameters are summarized in Table I.

To gain more insights into the quantum oscillations which are imitatively related to the FS of the $BaSn_3$ superconductor, we performed first-principle calculations on the electronic structure of $BaSn_3$ by the density functional theory (DFT). The band structure of $BaSn_3$ in the absence of spin-orbit coupling (SOC) is shown in Fig. 4(a), revealing a metallic behavior with several bands crossing the $E_F$ and the multi-orbital nature of the superconductivity. When SOC is taken into account, we focus our special attention on the bands near the $E_F$ along the Γ-A and K-H high symmetry lines.

Along the Γ-A direction, seen in Fig. 4(b), the lowest conduction band at the Γ point with $\Gamma_7^-$ irreducible representation (IR) shows a linearly downward dispersion and crosses with the first and second highest valence bands which belong to $\Gamma_{79}^-$ and $\Gamma_8^+$ IRs, respectively. The two unavoidable band crossing points due to the different IRs of relevant bands, see in Fig. 4(b), are labeled by red and blue filled circles and donated as DP1 and DP2, respectively. Due to protection by both time-reversal and space-inversion symmetries, each band in BaSn$_3$ is doubly degenerate, thus making DP1 and DP2 are actually fourfold degenerate type-I DPs. Figs. 4(c) and (d) present the band structure in the $k_x$ - $k_y$ plane surrounding DP1 and DP2, in which the band dispersion around both DP1 and DP2 is actually anisotropic, consistent with our above analysis on the dHvA oscillations. In Fig. 4(e), we show the electronic bands along the K-H direction, supporting the presence of type-II Dirac point, which is labeled by green filled circle and donated as DP3. Similarly, symmetry analysis shows that the DP3 is also unavoidable because of the crossing bands belong to different IRs ($\Lambda_4$ and $\Lambda_{5+6}$). The Dirac cone is tilted strongly along the K-H direction, whereas untilted in the $k_x$ - $k_y$ plane, seen in Fig. 4(f). Fig. 4(g) shows the schematic position of all the DPs in the BZ.

To further confirm the topological nature of BaSn$_3$ and examine the surface states, we plot the (001) surface band structure in Fig. 4(h). On the (001) surface, the bulk DP1 and DP2 located on the $k_z$ axis are projected onto the surface $\bar{\Gamma}$ point and hidden in the continuous bulk states. The surface states indicated by green arrows around the $\bar{\Gamma}$ point stemming from the projections of the bulk DP1 and DP2 are only partially observed. These topological states are far below the $E_F$, they consequently will not play a role in the measured transport properties, similar to the situation occurs in type-II DSMs PdTe$_2$ [63] and PtSe$_2$ [64]. In the case of DP3 that distributed on the K-H line, it is projected onto the surface $\bar{K}$ point. Though DP3 is 0.13 eV above the $E_F$, the surface state emanates from the projections of DP3 passes through the $E_F$ around the $\bar{K}$ point. This will contribute to the transport properties and help us with understanding the measured quantum oscillations in BaSn$_3$.

## IV. SUMMARY

As a summary, the de Hass-van Alphen quantum oscillations and first-principle calculations have demonstrated the coexistence of both type-I and type-II Dirac fermions in BaSn$_3$ superconductor. The magnetic quantum oscillations are originated from the type-II Dirac points formed along the high symmetric K-H direction of the Brillouin zone. The coexistence of type-I and type-II Dirac fermions makes BaSn$_3$ an excellent platform for the study of novel topological physics and exploring multiple functionalities. Moreover, since BaSn$_3$ is an intrinsic superconductor with nontrivial topological states, it would be very interesting to explore topological superconductivity which might be produced by the interplay between the two states. This holds possibility for making BaSn$_3$ more attractive for topological applications.

**ACKNOWLEDEMENTS**

The authors acknowledge the support by the Natural Science Foundation of Shanghai (Grant No. 17ZR1443300), the National Natural Science Foundation of China (Grant No. 11874264) and the strategic Priority Research Program of Chinese Academy of Sciences (Grant No. XDA18000000). Y.F.G. acknowledges the starting grant of ShanghaiTech University and the Program for Professor of Special Appointment (Shanghai Eastern Scholar) and the strategic Priority Research Program of Chinese Academy of Sciences (Grant No. XDA18000000). W.Z. is supported by the Shenzhen Peacock Team Plan (Grant No. KQTD20170809110344233), and Bureau of Industry and Information Technology of Shenzhen through the Graphene Manufacturing Innovation Center (Grant No. 201901161514). L.M.C thanks the Key Scientific Research Projects of Higher Institutions in Henan Province (19A140018).

**Table I.** Parameters derived from dHvA oscillations for BaSn$_3$.

| | $F$ (T) | $A_F$ (nm$^{-2}$) | $k_F$ (nm$^{-1}$) | $v_F$ (m/s) | $E_F$ (meV) | $m^*/m_0$ | $T_D$ (K) | $\tau$ (s) | $\mu$ (cm$^2$/Vs) | Berry phase |
|---|---|---|---|---|---|---|---|---|---|---|
| $B \perp (100)$ | 31.5 | 0.3 | 0.31 | $2.6 \times 10^5$ | 54 | 0.135 | 12 | $1 \times 10^{-13}$ | 1302 | 1.05($\delta = +1/8$) |
| | 196.6 | 1.87 | 0.77 | $5.03 \times 10^5$ | 258 | 0.178 | 3 | $4 \times 10^{-13}$ | 3951 | 0.17($\delta = +1/8$) |
| | 306.7 | 2.93 | 0.97 | $3.86 \times 10^5$ | 248 | 0.29 | 2.2 | $5.5 \times 10^{-13}$ | 3307 | -0.45($\delta = +1/8$) |
| $B \perp (001)$ | 300.8 | 2.87 | 0.96 | $8.03 \times 10^5$ | 509 | 0.138 | 18 | $6.7 \times 10^{-13}$ | 849 | -0.09($\delta = -1/8$) |

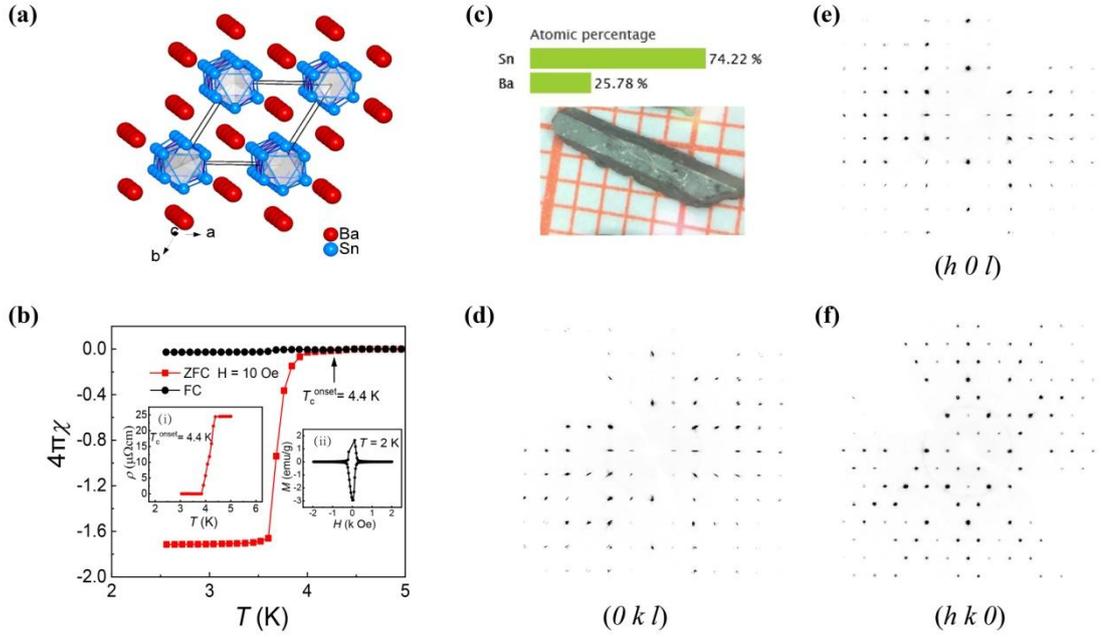

**Fig. 1 (Color Online).** (a) Schematic crystal structure of BaSn$_3$. (b) Temperature dependence of magnetic susceptibility. Inset (i) shows the temperature dependent resistivity of BaSn$_3$ crystal without application of magnetic field. Inset (ii) depicts the isothermal magnetization at 2 K, which displays a clear loop characterizing a type-II superconductor. (c) The stoichiometry of BaSn$_3$ crystal measured by the EDS spectrum and a picture of a typical crystal. (d)-(f) Diffraction patterns in the reciprocal space along the ($0\ k\ l$), ($h\ 0\ l$), and ($h\ k\ 0$) directions.

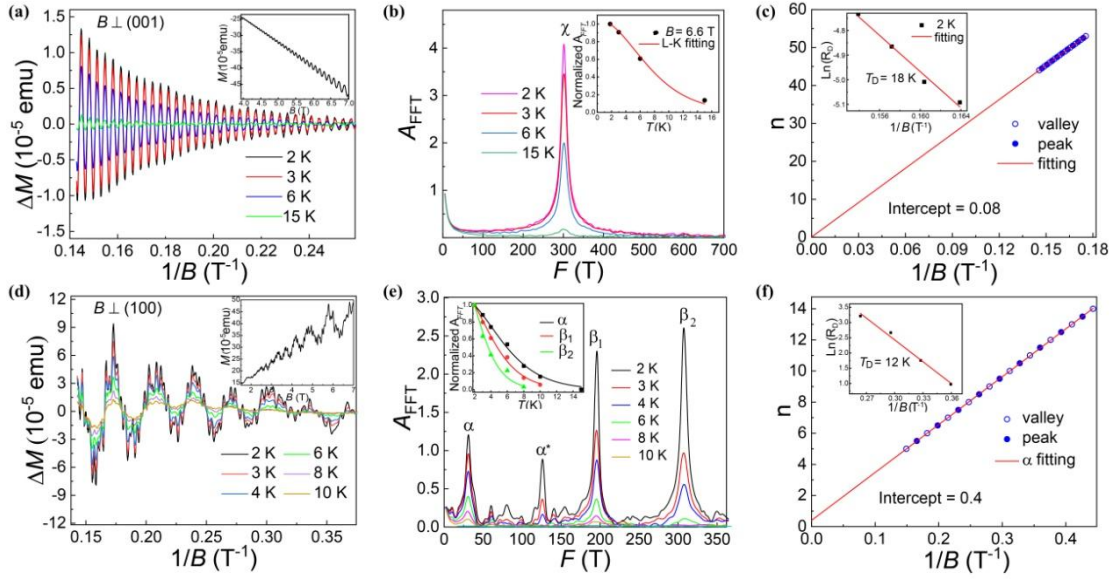

**Fig. 2 (Color Online).** (a) and (d) Quantum oscillations after subtracting the background for BaSn$_3$ with $B \perp (001)$ and $B \perp (100)$ at various temperatures. Each corresponding inset shows raw magnetization data at 2 K. (b) and (e) are temperature dependence of FFT spectra. Corresponding insets show the L-K fitting to the amplitudes of the dHvA oscillations. (c) and (f) The landau level fan diagrams for the fundamental frequencies of χ and α, respectively. Corresponding insets show the Dingle fitting results.

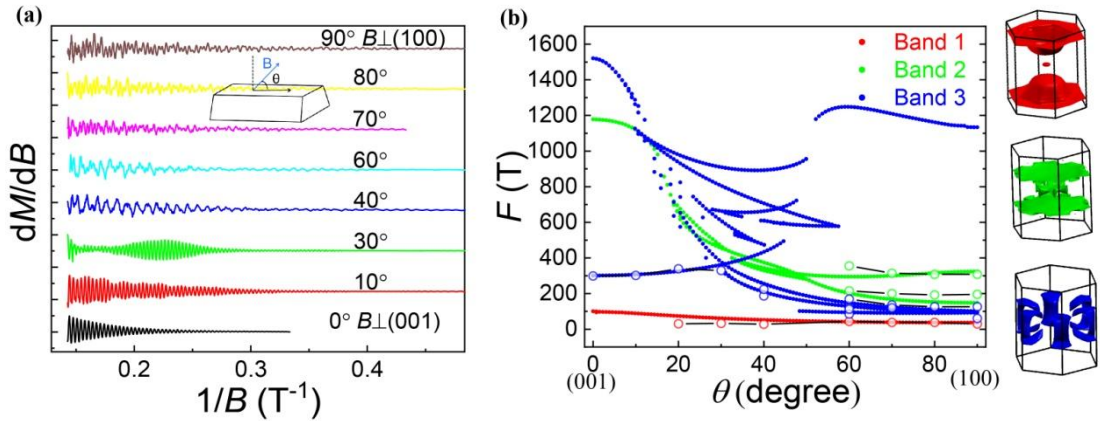

**Fig. 3 (Color Online).** (a) First order differential of dHvA oscillatory at $T = 2$ K for both $B \perp (100)$ and (001) for BaSn$_3$. Inset: the schematic measurement configuration. (b) Quantum

oscillatory frequencies as a function of angle θ. The solid dots depict the results of calculations and the color circles represent the experimental values. The corresponding FSs within the hexagonal first Brillouin zone are shown by the right figure.

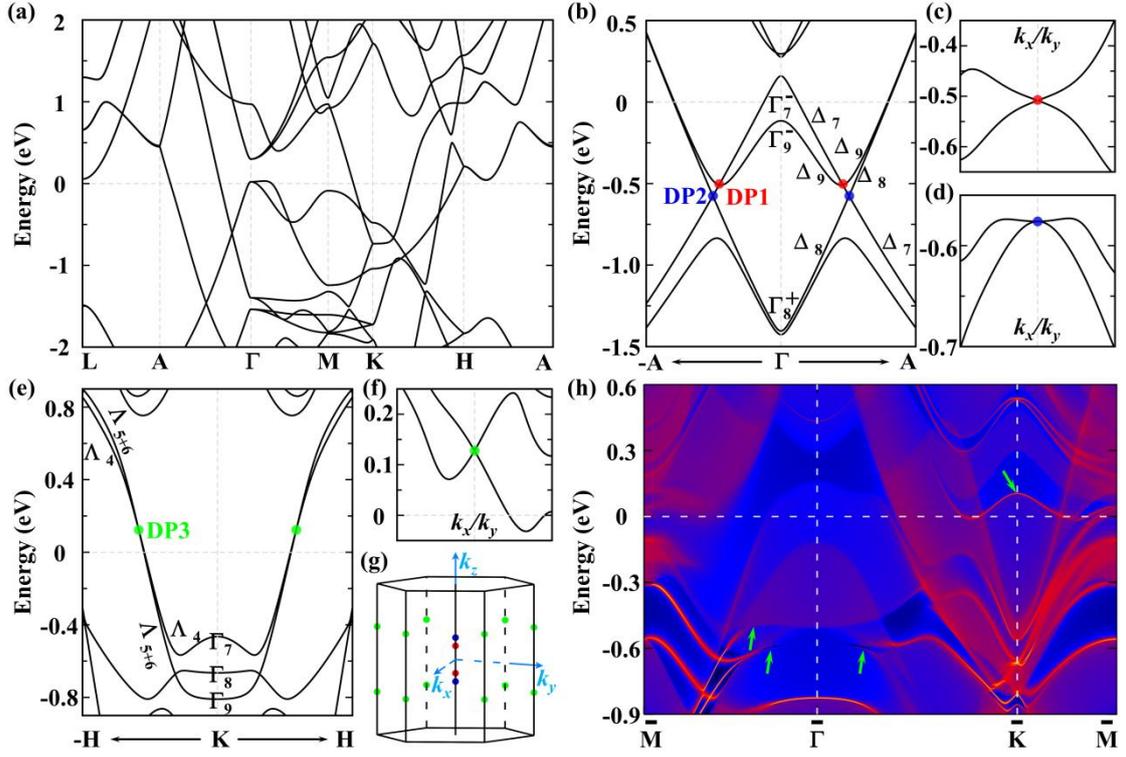

**Fig. 4 (Color Online).** (a) The calculated electronic band structure of $BaSn_3$ in the absence of SOC. (b) Energy dispersion along the Γ-A line in the presence of SOC, displaying two pairs of DPs. The irreducible representations of selected bands along the high-symmetric **k** are indicated. (c) and (d) Band structures in the $k_x$ - $k_y$ plane surrounding the DP1 and DP2 in (b). (e) Same as (b) but along the K-H line, showing a pair of DPs. (f) Band structure in the $k_x$ - $k_y$ plane in the vicinity of DP3 in (e). (g) Schematic position of the DPs in the BZ. (h) Topological surface states of $BaSn_3$ on the (001) surface. The green arrows mark surface states around the projected DPs.